%Paper: hep-th/9411122
%From: rball@surya11.cern.ch (Richard Ball)
%Date: Wed, 16 Nov 94 18:51:57 +0100
%Date (revised): Wed, 16 Nov 94 19:10:01 +0100
%Date (revised): Mon, 21 Nov 94 13:32:27 +0100
%Date (revised): Thu, 8 Dec 94 13:52:19 +0100

%%%%%%%%%%%%%%%%%%%%%%%%%%%%%%% v8.tex %%%%%%%%%%%%%%%%%%%%%%%%%%%%%%%%%%%%%%
%%%%%%%   Scheme Independence and the Exact Renormalization Group      %%%%%%
%%%%%%%    R.D. Ball, P.E. Haagensen, J.I. Latorre and E. Moreno       %%%%%%
%%%%%%%         in plain TeX with the harvmac macro package            %%%%%%
%%%%%%%     2 figures in compressed postscript in separate file        %%%%%%
%%%%%%%                  to be printed separately                      %%%%%%
%%%%%%%%%%%%%%%%%%%%%%%%%%%%%%%%%%%%%%%%%%%%%%%%%%%%%%%%%%%%%%%%%%%%%%%%%%%%%

\input harvmac
%\draftmode
\noblackbox

\pageno=0\nopagenumbers\tolerance=10000\hfuzz=5pt
\line{\hfill CERN-TH.7482/94}
\line{\hfill UB-ECM-PF 94/32}
\line{\hfill McGill/94-54}
\vskip 20pt
\centerline{\bf SCHEME INDEPENDENCE}
\centerline{\bf AND THE EXACT RENORMALIZATION GROUP
\footnote{$^*$} {This work is supported in part by funds provided
by Grant AEN 90-0033 (Spain), and by the M.E.C. (Spain).}}
\vskip 20pt\centerline{Richard~D.~Ball,$^{a}$\footnote{$^\dagger$}{On leave
 from a Royal Society University Research Fellowship.}
Peter~E.~Haagensen,$^{b}$
Jos\'e~I.~Latorre,$^{c}$ and Enrique~Moreno$^{c}$}
\vskip 12pt
\centerline{\it Theory Division, CERN,}
\centerline{\it CH-1211 Gen\`eve 23, Switzerland,\footnote{$^{a}$}
{E-mail: {\tt rball@surya11.cern.ch}}}
\vskip 8pt
\centerline{\it Physics Department, McGill University}
\centerline{\it 3600 University St., Montr\'eal H3A 2T8, Canada,
\footnote{$^b$}{E-mail: {\tt haagense@cinelli.physics.mcgill.ca}}}
\vskip 8pt
\centerline{\it Departament d'Estructura i Constituents de la Mat\`eria,}
\centerline{\it Facultat de F\'\i sica, Universitat de Barcelona,}
\centerline{\it Diagonal~647,~08028 Barcelona, Spain.\footnote{$^{c}$}
{E-mail: {\tt latorre@ecm.ub.es, moreno@ecm.ub.es}}}
\vskip 20pt
{\medskip\narrower
\ninepoint\baselineskip=9pt plus 2pt minus 1pt
\lineskiplimit=1pt \lineskip=2pt
\def\smallfrac#1#2{\hbox{${{#1}\over {#2}}$}}
\centerline{\bf Abstract}
\noindent
We compute critical exponents in a $Z_2$ symmetric scalar field theory in three
dimensions, using Wilson's exact renormalization group equations expanded in
powers of derivatives. A nontrivial relation between these exponents
is confirmed explicitly at the first two orders in the derivative
expansion. At leading order all our results are cutoff independent,
while at next-to-leading order they are not, and the determination of
critical exponents becomes ambiguous. We discuss the possible ways in
which this scheme ambiguity might be resolved.
}
\vskip 16pt
%\centerline{Submitted to: {\it Physics Letters B}}
\vskip 20pt
\line{November 1994\hfill}

\vfill\eject
\footline={\hss\tenrm\folio\hss}

%###########################################################################%

\def\ERG{exact renormalization group}

  \def\g{\gamma} 

 \def\l{\lambda} \def\L{\Lambda} \def\m{\mu}

   \def\vf{\varphi}

\def\pa{\partial}   \def\half{{1\over
2}}

 \def\frac#1#2{{{#1}\over {#2}}}
\def\half{\hbox{${1\over 2}$}}
 \def\smallfrac#1#2{\hbox{${{#1}\over
{#2}}$}}

\catcode`@=11 %This allows us to modify plain macros
\def\slash#1{\mathord{\mathpalette\c@ncel#1}}
 \def\c@ncel#1#2{\ooalign{$\hfil#1\mkern1mu/\hfil$\crcr$#1#2$}}
\def\lsim{\mathrel{\mathpalette\@versim<}}
\def\gsim{\mathrel{\mathpalette\@versim>}}
 \def\@versim#1#2{\lower0.2ex\vbox{\baselineskip\z@skip\lineskip\z@skip
       \lineskiplimit\z@\ialign{$\m@th#1\hfil##$\crcr#2\crcr\sim\crcr}}}
\catcode`@=12 %at signs are no longer letters

\def\PR{{\it Phys.~Rev.~}}

\def\NP{{\it Nucl.~Phys.~}}

\def\PL{{\it Phys.~Lett.~}}
\def\PRep{{\it Phys.~Rep.~}}
\def\AP{{\it Ann.~Phys.~}}

\def\RMP{{\it Rev.~Mod.~Phys.~}}
\def\IJMP{{\it Int.~Jour.~Mod.~Phys.~}}

\def\ZP{{\it Zeit.~Phys.~}}

\def\vol#1{{\bf #1}}\def\vyp#1#2#3{\vol{#1} (#2) #3}

\nref\wilson{K.~Wilson and J.~Kogut, \PRep.\vyp{12}{1974}{75}\semi
             K.~Wilson, \RMP\vyp{47}{1975}{773}.}
\nref\wh{F.J.~Wegner and A.~Houghton, \PR\vyp{A8}{1972}{401}.}
\nref\weinberg{S.~Weinberg, in ``Understanding the Fundamental
Constituents of Matter'', Erice~1976, ed.~A.~Zichichi
(Plenum,~1978).}
\nref\polch{J.~Polchinski, \NP\vyp{B231}{1984}{269}.}
\nref\hh{A.~Hasenfratz and P.~Hasenfratz, \NP\vyp{B270}{1986}{687}.}
\nref\margaritis{A.~Margaritis, G.~Odor and A.~Patkos,
\ZP\vyp{C39}{1988}{109}.}
\nref\hklm{P.E.~Haagensen, Yu.~Kubyshin, J.I.~Latorre and E.~Moreno,
\PL\vyp{323B}{1994}{330}.}
\nref\timzero{T.R.~Morris, \IJMP\vyp{A9}{1994}{2411}.}
\nref\alford{M.~Alford, \PL\vyp{336B}{1994}{237}.}
\nref\rusoone{A.E.~Filippov and S.A.~Breus, \PL\vyp{A158}{1991}{300}.}
\nref\golner{G.R.~Golner, \PR\vyp{B33}{1986}{7863}.}
\nref\rusotwo{A.E.~Filippov and A.V.~Radievsky, \PL\vyp{A169}{1992}{195}.}
\nref\timone{T.R.~Morris, \PL\vyp{329B}{1994}{241}.}
\nref\tetwet{N.~Tetradis and C.~Wetterich, \NP\vyp{B422}{1994}{541}.}
\nref\bt{R.D.~Ball and R.S.~Thorne, \AP\vyp{164}{1994}{117}.}
\nref\timtwo{T.R.~Morris, \PL\vyp{334B}{1994}{355}.}
\nref\rusozero{V.I.~Tokar, \PL\vyp{A104}{1984}{135}.}

%\newsec{Introduction}\bigskip

The exact renormalization group \refs{\wilson-\polch} is in many
instances a powerful analytical technique for the study of
nonperturbative renormalization group flows.  In practice one must
always truncate an infinite tower of equations, and in the past
surprisingly good results for critical exponents and renormalization
group flows of scalar field theories in $2\le d\le 4$ have been obtained
\refs{\hh-\rusoone} by suppressing all the equations except that for the
effective potential.  There have been several attempts
\refs{\golner-\tetwet} to improve the accuracy of these calculations by
also keeping terms in the effective action containing two derivatives of
the fields, and thus allowing for wave function renormalization.  In
this letter we will consider the important issue of scheme dependence
within this context.

A Euclidean
field theory may in general be regularized by employing a certain
cutoff function $K_{\L}(p)$, which suppresses modes with
momenta higher than some scale $\L$.  The \ERG\ equations are then
constructed so as to ensure that as $\L$ is reduced, and thus more and
more modes are integrated out, the vertex functions of the theory are
varied in such a way that the Green's functions
remain unchanged.  This provides a nonperturbative definition of the
theory.  Now in principle physical observables should be determined
unambiguously for a broad class of cutoff functions $K_{\L}(p)$.  This
is what we mean by scheme independence.  Indeed in certain truncation
schemes this independence can be proven; for instance, it can be shown
that S-matrix elements are scheme independent to any given order in
perturbation theory\bt, even though beta-functions and anomalous
dimensions are in general scheme dependent beyond leading order.
However in other truncations scheme independence cannot be taken for
granted.  In the derivative expansion, in particular, this problem has
not so far been considered: a particular choice of cutoff function is made
(e.g., sharp cutoffs \refs{\hh-\alford}, power-like cutoffs \timone, or
exponential cutoffs \refs{\rusotwo,\tetwet}), giving perhaps the impression
that the results obtained are in some sense unique.

Here we will compute various critical exponents for a $Z_2$-symmetric scalar
field theory in $d\!=\!3$ dimensions using the simple and elegant \ERG\
equations due to Wilson \wilson\ and Polchinski \polch: previous
authors \refs{\hh-\alford,\timone,\tetwet} have based their computations
on the Wegner-Houghton \wh\ and Weinberg
\weinberg\ equations. We work to all orders in the fields, but
expand the (Wilson) effective action in powers of derivatives. We will
show that all the scheme dependence can be absorbed into $2n$
parameters at $n$-th order in the expansion.  At leading order, $n=0$,
the critical exponents are then scheme independent: they are
independent of the particular choice of cutoff function.  However
at next-to-leading order the results become scheme dependent. This makes it
particularly difficult to compute unambiguously the critical exponent
$\eta$ (corresponding to wave function renormalization). We will
attempt to control this dependence by employing a minimal sensitivity
criterion.

We also discuss the `scaling' relation between the magnetic
deformation exponent $\lambda_H$ and $\eta$,
showing explicitly that this is satisfied exactly (in any scheme) at
the first two orders in the derivative expansion.

\newsec{The Exact Renormalization Group}

Exact renormalization group equations are normally obtained in two
steps: incomplete integration of modes, followed by rescaling.  The
first step is most easily described in the path integral approach
developed by Polchinski \polch.  Consider the following Wilson action for a
scalar field theory
\eqn\action{S[\varphi;\Lambda]\equiv\half\int_p\varphi_p\varphi_{-p}
P_{\Lambda}^{-1}(p^2) + S_{\rm int}[\varphi;\Lambda],} where
$\int_p\equiv\int\smallfrac{d^{d}p}{(2\pi)^d}$, $P_{\Lambda}(p^2)$ is an
analytic function \bt\ with a single pole on the negative real axis of
unit residue, which falls sufficiently fast as $p^2/\Lambda^2\to\infty$
that all modes
with Euclidean momenta $p^2\gg\Lambda^2$ are suppressed, and $S_{\rm
int}$ contains all higher order interactions.  We will write
generically, \eqn\cutoff{P_{\Lambda}(p^2)={K(p^2/\L^2)\over p^2}\ ,}
where $K(p^2/\L^2)$ is the cutoff function responsible for damping high
momentum modes. The \ERG\ then makes use of two momentum scales,
an ultraviolet regulator scale $\Lambda_0$,
which is held fixed, and $\Lambda$ which controls the scale down to
which high momentum modes have been (loosely speaking) integrated
out.\foot{More properly we
should also include a mass scale $m\ll\Lambda_0$ (the position of the
zero in $P_\Lambda$); while important in the extreme infrared
($\Lambda\ll m$), it may be ignored in the vicinity of the Wilson
fixed point, since there $m/\Lambda$ will be very small.}
What is of relevance for the flow is their ratio or, equivalently,
$t\equiv\ln {\Lambda_0\over \Lambda}$; renormalization group flows then
go from the far ultraviolet ($t=0, \Lambda=\Lambda_0$) to the far infrared
($t\rightarrow \infty, \Lambda\rightarrow 0$).
As $\Lambda$ is reduced from $\Lambda_0$ to zero the interaction,
$S_{\rm int}[\phi;\Lambda]$, must be
evolved in such a way that the amputated Green's functions of the theory
remain unaltered; this can be achieved if $S_{\rm int}$ satisfies
the renormalization group equation
\eqn\ergeqpol{\Lambda\frac{d}{d\Lambda}S_{\rm int}= \half
\int_p\Lambda\frac{\pa P_\Lambda(p^2)}{\pa\L}\left( {\delta S_{\rm
int}\over \delta \varphi_p} {\delta S_{\rm int}\over \delta
\varphi_{-p}}
- {\delta^2 S_{\rm int}\over \delta \varphi_p\delta
\varphi_{-p}}\right)\ - \half \eta(t) \int_p
\varphi_p\varphi_{-p} P_{\Lambda}^{-1}(p^2).}
(Contact terms have been suppressed here; see ref.\bt\ for more
details.) Here $\eta(t)$ is the anomalous dimension of the
(renormalized) field $\varphi_p$:
\eqn\eeta{\Lambda\frac{d}{d\Lambda} \varphi_p = -\half \eta(t)
\varphi_p.}

The above renormalization group equation \ergeqpol\ may be further
simplified by rewriting it in terms of the complete Wilson action
$S[\varphi;\Lambda]$: suppressing the contact term this gives
\eqn\ergeq{\Lambda\frac{d}{d\Lambda}S=
 \half \int_p\Lambda\frac{d}{d\Lambda}P_\Lambda(p^2) \left(
{\delta S \over \delta \varphi_p}
{\delta S \over \delta \varphi_{-p}} -
{\delta^2 S \over \delta \varphi_p\delta \varphi_{-p}}
-2P_\Lambda^{-1}(p^2)\varphi_p{\delta S \over \delta
\varphi_p}\right).}
The explicit $\eta$ dependence has now dropped out because
renormalizations of the fields can be absorbed directly into
corresponding renormalizations of the vertices.
As we shall see, this equation \wilson\ is rather easier to use
than the other \ERG\ equations \refs{\wh,\weinberg} employed in
\refs{\hh-\alford,\timone,\tetwet}, due essentially to the
relative simplicity of its non-linear term.

Finally we consider the rescaling step. For this purpose we have to
write the couplings in $S$ in
a manifestly dimensionless way, since it is dimensionless couplings
which parametrize the theory and whose flow we should study.
Thus, we write
\eqn\sint{S_{\rm int}
=\sum_{n={\rm even}}\int_{\hat p_1\dots\hat p_n}
s_n(\hat p_1,\dots,\hat p_n;t)\
\hat\varphi_{p_1} \cdots \hat\varphi_{p_n}
\ \delta(\hat p_1+\dots +\hat  p_n)\ ,}
where $\hat p_i\equiv \Lambda^{-1}p_i$,
$\hat\varphi_p\equiv\Lambda^{1+d/2}\varphi_p$, and the couplings (or
`vertex functions') $s_n(\hat p_1,\dots,\hat p_n;t)$
are dimensionless. We can then rewrite \ergeqpol\ as a set of
equations for the flow of the couplings $s_n$:
\eqn\finalerg{\eqalign{
\dot S=&\int_{\hat p} K'({\hat p^2})
\left({\delta S\over \delta \hat\varphi_p}
      {\delta S\over \delta \hat\varphi_{-p}} -
      {\delta^2 S\over \delta \hat\varphi_p\delta\hat\varphi_{-p}} \right)
 + d\ S\cr
&+\int_{\hat p} \left(1-\smallfrac{d}{2}-\smallfrac{\eta(t)}{2}
-2\hat p^2 \frac{K'(\hat p^2)}
{K(\hat p^2)}\right)
\hat\varphi_p \ {\delta S\over \delta\hat\varphi_p} -
\int_{\hat p} \hat\varphi_p\ \hat p\cdot{\partial'\over \partial\hat p}
{\delta S\over \delta\hat\varphi_p}\ ,\cr}}
where $\dot S$ is given by a similar expression to \sint, but
with the dimensionless couplings $s_n$ replaced by their partial
derivatives with respect to $t$, and
the prime in the momentum derivative means that it does not act
on the momentum conserving delta functions of \sint .

The form of $\eta(t)$ along the flow is not yet determined: it
will depend on the chosen normalization of the field $\varphi_p$.
For the bare (unrenormalized) fields used in \refs{\polch,\bt}
$\eta(t)=0$, and the interaction
$S_{\rm int}$ will develop quadratic terms which change the
residue of the pole in the propagator.
Alternatively one can choose the anomalous dimension $\eta(t)$
in such a way that no such terms arise, and the residue thus remains
fixed, by imposing the normalization condition
\eqn\residue{ \Lambda\frac{d}{d\Lambda} \left(
{\partial \over \partial p^2} \left. {\delta^2 S\over
\delta \varphi_p \varphi_{-p}}\right|_{{p^2=0} \atop
\varphi=0}\right)=0,}
at each point of the renormalization group flow. This then
implicitly defines the anomalous dimension $\eta(t)$, which at
fixed points of the flow becomes equal to the critical
exponent $\eta_*$.

\newsec{Derivative Expansion and Scheme Dependence}

The overwhelming complexity of the full \ERG\ equations means that in
practice some sort of truncation must be employed.  Since the action may
be naturally expanded in powers of the fields and their derivatives, it
is natural to consider truncations of either or both of these
expansions.  Truncations in the number of powers of the fields, as
employed in \refs{\margaritis-\timzero} seem to be poorly convergent,
with many spurious solutions (see \timtwo\ for a detailed analysis),
although the convergence seems to improve if the expansion is made about
the minimum of the potential \refs{\alford,\tetwet}.  A better procedure
\refs{\hh,\golner-\timone} seems to be to work to all orders in the
number of fields, and expand the interaction only in powers of
derivatives.  In this way it is hoped that the essential features of the
long-wavelength physics will be retained.

Truncating at second order in derivatives, we thus write
\eqn\expansion{
S_{\rm int}=\Lambda^d\int d^d\!x \left( v(\hat\vf ,t) + z(\hat\vf ,t)
\Lambda^{-2}\big(\pa_\m \hat\vf \big)^2 +\cdots \right)}
or, alternatively,
\eqn\altexp{s_n(\hat p_1,\dots,\hat p_n;t)=v_n(t)+\smallfrac{1}{n(n-1)}
{(\hat p_1^2+\dots +\hat p_n^2)}\ z_{n-2}(t)+\cdots \ ,}
with $v(\hat\vf ,t)=\sum v_n(t)\hat\vf^n$,
$z(\hat\vf ,t)=\sum z_n(t)\hat\vf^n$, and
where now $z_0(t)=0$ due to the normalization condition \residue.
The \ERG\ equation
\finalerg\ can now be projected onto a series of equations, order by
order in $p^2$, in which $\hat\varphi$ is taken to be effectively
a constant (and denoted by $x$). This
procedure is somewhat lengthy but unambiguous, and results
in an infinite set of coupled
partial differential equations:\foot{The zeroth order equation (with
$z=\eta=0$) seems to have been first derived in \rusozero; different
versions of the first order equations may be found in
\refs{\golner,\rusotwo}.}
\eqn\eqset{\eqalign{
\dot v&= I_0 v''+2 I_1 z- K_0 v'^2 +
(1-\smallfrac{d}{2}-\smallfrac{\eta}{2}) x v' + dv\ ,\cr
\dot z&= I_0 z''+K_1 v''^2-4 K_0 z v'' -2 K_0 z' v'+
(1-\smallfrac{d}{2}-\smallfrac{\eta}{2}) x z'-\eta z
-\smallfrac{\eta}{2}\ ,\cr
&\vdots\cr}}
where dots and primes denote partial derivatives with respect to $t$
and $x$ respectively. The number of equations grows rather fast with
increasing powers of $p^2$;
there are three more equations at the next order. So far,
there seems to have been no attempt to go to order $p^4$; here
we will just study the two equations \eqset.
The numbers $K_n$, $I_n$, $n=0,1,\ldots$
are remnants of the cutoff function $K(z)$ and
thus parametrize the scheme dependence of the equations:
\eqn\constants{K_n\equiv (-)^{n+1}K^{(n+1)}(0),\qquad
I_n\equiv -\int_{\hat p} ({\hat p}^2)^n K'({\hat p}^2)
=-\Omega_d\int_0^\infty\!dz z^{d/2-1+n}K'(z)\ ,}
where $K^{(n)}$ is the $n$-th derivative of $K$, and
$\Omega_d=2/\big(\Gamma(\smallfrac{d}{2})(4\pi)^d\big)$.
At zeroth order in $p^2$, all the cutoff dependence
is parametrized in terms of $K_0$ and $I_0$, while at first
order $K_1$ and $I_1$ are also needed; at $n$-th order we need $K_n$ and
$I_n$. Thus the further we go in the derivative
expansion, the more information about the cutoff function is
needed; short-distance properties of it only gain relevance as the
derivative expansion is carried to higher orders. Incidentally,
we note that in a calculation of perturbative renormalization group
functions in this setting, these same remnants of the cutoff
function are present and determine the scheme dependence of some of the
results\ref\hughes{J. Hughes and J. Liu, \NP\vyp{B307}{1988}{183}.}.

We have just shown that the cutoff dependence of eq. \eqset\ is given in
terms of at most $2n+2$ arbitrary parameters at order $p^{2n}$.  In
fact we can exploit global rescalings of the
variable $x$ and of the function $v$ to further simplify the
first two equations to
\eqn\finaleqset{\eqalign{ \dot v&= v''+2 A z- v'^2 +
(1-\smallfrac{d}{2}-\smallfrac{\eta}{2}) x v' + dv\ ,\cr
\dot z&= z''+B v''^2-4 z v'' -2  z' v'+
(1-\smallfrac{d}{2}-\smallfrac{\eta}{2}) x z'-\eta z
-\smallfrac{\eta}{2}\ ,\cr}} where
\eqn\alphabeta{ A\equiv{I_1 K_0\over
I_0},\qquad B\equiv{K_1\over K_0^2}\ .}
It follows that at zeroth order ($z=0$) there is no cutoff
dependence at all, while all cutoff dependence at order $p^2$ is reduced to a
two-parameter family $(A,B)$.  Furthermore, reasonable cutoff functions
(for example, exponentials of polynomials\bt) will decrease smoothly and
monotonically, so all moments $I_n$ will be positive and of similar
magnitude; the derivatives at the origin, $K_n$, will also be loosely of
the same order.  In other words, $A$ and $B$ will both be positive and
loosely of order unity.  For example, with an exponential cutoff
\refs{\rusotwo,\tetwet} $K(z)=e^{-z}$, $A = d/2$ and $B = 1$.  On the
other hand a sharp cutoff $K(z)=\theta(1-z)$, corresponds to a singular
point in the $A$-$B$ parameter space, since $A = 0$ while $B$ is
undetermined, and can thus only be made well-defined through some
limiting procedure (as discussed in \timzero).  The power-like cut-off
$K(z)=(1+z^2)\inv$ employed in \timone\ is also singular since $A=0$ and
$B$ is infinite.  In these cases it becomes necessary to use other
more complicated \ERG\ equations
\refs{\wh-\weinberg}.

\newsec{Leading Order Numerical Results}

The truncation of the derivative expansion of the \ERG\ equation at
zeroth order is obtained trivially from eq. \finaleqset\ by writing
$z(x,t)=0$ and $\eta=0$.  Setting $\eta$ to zero is
justifed only by the a priori suppression of all momentum
dependent terms in the interaction; it turns out to be a good
approximation at zeroth order because the actual value
of $\eta_*$ is very small.
Writing $f\equiv v'$ (to eliminate the arbitrary field independent
part of $v$) the leading order \ERG\ equation is then
\eqn\zerothorder{\dot f=f'' - 2 f
f'+(1+\smallfrac{d}{2}) f +(1-\smallfrac{d}{2}) xf' \ .} This is
considerably simpler than equivalent truncations of other
\ERG\ equations\refs{\wh,\weinberg}, since all non-linearity is
reduced to a quadratic term which is easily controlled numerically.
Furthermore, eq. \zerothorder\ is manifestly scheme independent (that
is, independent of the particular choice of cutoff function).  This
is of great importance since the results we will display shortly are the
leading orders of the expansion.

We begin by solving the fixed point equation \eqn\zerothfixedpoint{ 0=
{f_*}''-2 f_* {f_*}' +(1+\smallfrac{d}{2}) f_*+ (1-\smallfrac{d}{2}) x
{f_*}'\ ,} with initial conditions $f_*(0)=0$ (by the $Z_2$ symmetry)
and ${f_*}'(0)=\gamma$.  We now concentrate for definiteness on
dimension $d=3$ where a comparison with other results is readily
available.  The nontrivial solution at the ultraviolet fixed point only
exists for one value of the initial condition $\gamma=\g_*$.
Numerically, this is found by tuning $\gamma$ so as to let $f_*$ run as
far as possible; in this way we determine
$\gamma_*=-.228601293102\dots\ $ (where all figures are significant:
the behaviour of $f_*(x)$ at large $x$ is extremely sensitive to the
precise value of $\g_*$, always diverging at some finite value of
$x$ for $\g\neq\g_*$).

It is in fact trivial to determine the asymptotic behavior of the
solution $f_*(x)$ to \zerothfixedpoint\ at large $x$: $f_*(x)\sim x$,
and thus $v_*(x)\sim x^2$.
This is in marked contrast to the $x^5$ asymptotic behavior of solutions
to the equations considered in
\refs{\hh,\timone}, resulting from their rescaling terms.  This
difference stems from the different structure of the non-linearities in
the equations; we find it intriguing since it is ultimately responsible
for the differences in the numerical results.

Departures from the fixed point solution are governed by operators whose
critical exponents are universal.  In our case, the eigenvalue equation
to be solved is given by the linearization of eq. \zerothorder: writing
$f(x,t)=f_*(x)+ \sum_n a_ng_n(x)e^{\lambda_n t}$, with $a_n$ small,
\eqn\zerotheigen{g_n'' -2 (f_* g_n'+{f_*}' g_n) +\smallfrac{5}{2} g_n-
\smallfrac{x}{2} g'_n= \lambda_n g_n\ ,} where $g_n$ is the $n$-th
eigenfunction with eigenvalue $\lambda_n$.  The critical exponent of the
only relevant, $Z_2$-symmetric, operator at the infrared fixed point is
known as $\nu\equiv 1/\lambda_1$, while the first irrelevant one is
known as $\omega\equiv -\lambda_2$.  We investigate eq. \zerotheigen\
numerically and find the results given in Table~III below (LO).  Further
eigenvalues for irrelevant operators may be determined similarly, but
become increasingly meaningless as they depend more and more on
short-distance dynamics. Curiously our leading order result
$\nu=0.649$ is rather closer to the values obtained using more
traditional methods
\ref\itdr{C.~Itzykson and J.M.~Drouffe, {\it Statistical Field Theory,
Vol. 1} (Cambridge University Press, 1989)\semi
J.~Zinn-Justin, {\it Quantum Field Theory and Critical Phenomena},
 (Oxford University Press, 1990).} than that obtained
using the leading order approximation to other \ERG\ equations
(i.e. $\nu = 0.690$ for a Wegner-Houghton equation
\refs{\hh-\alford}, $\nu = 0.660$ for an equation for one particle
irreducible vertices with a power cut-off \timone). However before we
can attach any significance to this observation we need to determine
the size of the error induced by the truncation of the derivative
expansion, and this we can only do by calculating the next order correction.

\newsec{Next-to-Leading Order}

The main drawback of the simple zeroth order approximation described
above is the absence of wave function renormalization: $\eta$ is set
to zero.
Equation \zerothorder\ thus includes only the physics of zero modes, whereas
wave function renormalization is precisely related to the kinetic term
which carries derivatives.  This leads us to consider the
next-to-leading corrections, of order $p^2$ in the derivative expansion.

To study numerically the order $p^2$ approximation of the exact
renormalization group equation eqns. \finaleqset\ we first consider the
fixed point differential equations ${\dot f}(t,x)=0,\ {\dot z}(t,x)=0$:
\eqn\finaleqsetfix{\eqalign{ &0 = f_*''+2 A z_*'-2 f_* f_*'+\Delta^+ f_*
+\Delta^- xf_*'\ ,\cr &0 = z_*''+B f_*'^2-4 z f_*' -2 z' f_*+\Delta^-
xz_*' -\eta z_*-\smallfrac{\eta}{2}\ ,\cr}} where $\Delta^{\pm}\equiv 1
\pm d/2 - \eta /2$.  This pair of non-linear coupled ordinary
differential equations, together with appropriate initial conditions,
determine the shape of the Wilson fixed point action at order $p^2$.
Three of the initial conditions are already fixed: two by invoking $Z_2$
symmetry, $f_*(0)=0$ and $z_*'(0)=0$, and the third by the normalization
condition eq. \residue, $z_*(0)=0$.  For any fixed values of the
parameters $A$ and $B$, a well-behaved solution of eqs. \finaleqset\
({\it i.e.}, a solution defined for all finite $x$) only exists for the
unique value $\eta=\eta_*$ of the anomalous dimension and a unique value
$\gamma_*$ of the initial condition $f_*'(0)=\gamma$.  For any other
choice of $\eta$ and $f_*'(0)$ the solutions $f_*(x)$ and $z_*(x)$ end
at singularities at finite $x$.  By contrast, the correct solutions
behave asymptotically, as $x \to \infty$, as \eqn\asym{\eqalign{ f_*(x)
&\sim (1-\eta_*/2)\ x + k x^{-\Delta^+ /\Delta^-} - 2 k^2
{\Delta^-\over\Delta^+(2-\eta_*)}\ x^{-2\Delta^+/\Delta^- -1} + O(x^{-3
\Delta^+ / \Delta^- -2}), \cr z_*(x) &\sim
\frac{B(1-\smallfrac{\eta_*}{2})^2-\smallfrac{\eta_*}{2}} {4 - \eta_*} +
k{\Delta^- \over \Delta^+ } \frac{B (\eta_*
-2)-\eta_*}{4-\eta_*}\ x^{-\Delta^+ /\Delta^- -1} +
O(x^{-2\Delta^+ /\Delta^- -2}),\cr}} where $k$ is a constant.

\topinsert\hfil
$$\vbox{\tabskip=0pt \offinterlineskip
\halign to350pt{\strut#& \vrule#\tabskip=1em plus2em&
   \hfil#& \vrule#& \hfil#& \vrule#&
   \hfil#& \vrule#&
   \hfil#& \vrule#\tabskip=0pt\cr
\noalign{\hrule}
& & \hidewidth$\Delta\eta\times 10^3$\hidewidth
& &\hidewidth $B=0.25$\hidewidth
& &\hidewidth $0.5$\hidewidth
& &\hidewidth $0.75$\hidewidth&
 \cr \noalign{\hrule}
 &&$A=0.5$&& -0.76 & & - 2.8& &  -6.1& \cr \noalign{\hrule}
 & & 0.7      & & -0.28 & & - 0.9& &  -1.9& \cr \noalign{\hrule}
 & & 0.8      & & -0.03 & &   0.0& &   0.2& \cr \noalign{\hrule}
 & & 0.9      & &  0.21 & &   1.0& &   2.5& \cr \noalign{\hrule}
 & & 1.1      & &  0.71 & &   3.0& &   7.1& \cr \noalign{\hrule}
\noalign{\smallskip}}}$$
\hfil
\centerline{\vbox{\hsize= 300pt \noindent\footnotefont
Table~I: The difference \hbox{$\Delta\eta\equiv\eta_2-\eta_1$} for a range
of values of the scheme parameters $A$ and $B$. }}
\bigskip
\endinsert

Equations \finaleqsetfix\ are nonlinear and stiff.  Furthermore,
finding the correct solution involves a double fine tuning in
$\eta$ and $f_*'(0)$.  Thus it becomes extremely difficult to simply
integrate up the equations as we did in the zeroth order case, and
it is much easier to proceed
recursively.  So first we set $z_*(x)$ and $\eta$ to zero and find the
unique value $\gamma_0$ of $f'(0)$ for which the first equation
\finaleqsetfix\ has a non-singular solution, $f_0(x)$, just as we did
for the zeroth order approximation.  Now we use this solution in the
second equation \finaleqsetfix\ and solve for $z_1(x)$ by fine-tuning
$\eta$ to the appropriate value $\eta_1$.  With the values of $\eta_1$
and $z_1(x)$ we return to the first equation and obtain a new solution
$f_1(x)$, with $f_1'(0)=\gamma_1$, and so on.  In this way we obtain a
sequence of functions $f_0,z_1,f_1,z_2,f_2,\ldots$ and constants
$\gamma_0,\eta_1,\gamma_1,\eta_2,\gamma_2,\ldots$ which (at least for a
reasonable range of $A$ and $B$) converge rather rapidly to the exact
solution $f_*(x),z_*(x),\gamma_*, \eta_*$.
Solutions $f_*(x)$ and $z_*(x)$ are shown in \fig\fandz{The fixed
point solution at next to leading order for $A=0.8$ and $B=0.25,0.5,0.75$:
a) $f_*(x)$ b) $z_*(x)$. The (scheme independent) leading order
solution is also shown (dashed).}.

Again for $d=3$, a numerical analysis of this sequence of solutions
shows that the rate of convergence of the iterative procedure is in
effect controlled by $A$.  Indeed for any fixed $B$, the convergence is
fastest (in the sense that corrections to $\eta$ in sucessive iterations
are minimized) when $A=0.8$.  Moreover, the function $\eta(A)$ has an
inflexion point at this particular value of $A$.  This is demonstrated
in Table I where we show the difference $\Delta\eta\equiv\eta_2-\eta_1$
for different values of the parameters $A$ and $B$.  We will use this
criterion to fix, from now on, the value of $A$ to $0.8$.

Within the accuracy of our computations, we could not find a similar
minimal sensitivity criterion for the parameter $B$ either in the fixed
point solutions or in the critical exponents. The values
of $\eta_*$ are given in Table II. The dependence of the
anomalous dimension $\eta_*$ in the
parameter $B$ is almost linear, as can be inferred by the following
argument.  By varying the initial conditions of $f_*$ and $z_*$ at $x=0$
in the first equation \finaleqset\ and assuming a (fast) convergence of
the iterative process, we can relate the second derivative of $z_1$ at
the origin to the parameter $\gamma_0$.  Using this relation and the
projection onto $x=0$ of the second equation \finaleqset\ one can deduce
the approximate relation: \eqn\appbeta{ \eta_*\approx 2 B {{\gamma^*}^2
\over 1- {\gamma^* \over A}}\ ,} in good qualitative agreement with our
numerical results.  Although no serious results can be obtained from
this estimate of $\eta_*$, it is interesting that it has no stationary
points in either $A$ or $B$.  The optimal value of $A=0.8$ is, thus, due
to higher order iterations.  However, to the precision we work with, the
dependence of $\eta_*$ on $B$ remains almost linear even when subsequent
iterations are considered.

\topinsert\hfil
$$\vbox{\tabskip=0pt \offinterlineskip

\halign to350pt{\strut#& \vrule#\tabskip=1em plus2em&
   \hfil#& \vrule#& \hfil#& \vrule#&
   \hfil#& \vrule#&
   \hfil#& \vrule#\tabskip=0pt\cr
\noalign{\hrule}
& &\hidewidth \hidewidth
& &\hidewidth $\eta$ \hidewidth
& &\hidewidth $\nu$ \hidewidth
& &\hidewidth $\omega$ \hidewidth &
 \cr \noalign{\hrule}
 & & B = 0.25 & & 0.01898 & & 0.63706 & & 0.7016
                                    & \cr\noalign{\hrule}
 & & 0.5  & & 0.03762 & & 0.62522 & & 0.7733
                                    & \cr\noalign{\hrule}
 & & 0.75 & & 0.05595 & & 0.61595 & & 0.8501
                                    &\cr\noalign{\hrule}
\noalign{\smallskip}}}$$
\hfil
\centerline{\vbox{\hsize= 300pt \noindent\footnotefont
Table~II: Critical exponents at next-to-leading order for
\hbox{$A=0.8$} and a range of values of B. }}
\bigskip
\endinsert

Once a fixed point solution is found we can study the critical
exponents associated with this solution, by again linearizing
equations \finaleqset\ around the fixed point.
This gives rise to the eigenvalue problem
\eqn\lineq{\eqalign{
g_n''(x)&-2 g_n(x) {f_*}'(x) -2 f_*(x) g_n'(x) + \Delta^+g_n(x) +
\Delta^- x g_n'(x) + 2 A h_n'(x)= \lambda_n g_n(x),\cr
h_n''(x)&-2 f_*(x) h_n'(x)-2 g_n(x) {z_*}'(x) - 4 {f_*}'(x) h_n(x) -
4 g_n'(x) z_*(x)\cr
& \phantom{aaaaaaaaaaaaaaa} +\Delta^- x h_n'(x)-
\eta_* h_n(x)+ 2 B {f_*}'(x) g_n'(x)= \lambda_n h_n(x) , \cr}
}
where $f_*(x)$ and $z_*(x)$ are the fixed point solutions, $\eta_*$
the corresponding anomalous dimension,
and $\left(g_n(x), h_n(x) \right)$ is the eigenvector
(scaling operator) corresponding to the critical exponent $\lambda_n$.

For $Z_2$-symmetric perturbations, we now use as initial
conditions $g_n(0)=0, h'_n(0)=0$, which are fixed by the $Z_2$-symmetry,
and $g'_n(0)=1$ as a normalization.  Solutions
are again found by an iterative procedure, where the two
parameters to be tuned are now $\l_n$ and $h_n(0)=h_0$.\foot{
Although $h_0$ is very small, it is nonzero due to the fact that we
have implicitly relaxed the normalization condition eq. \residue\ by
setting  $\eta(t)=\eta_*$ away from the fixed point.}
When $d=3$, we find as expected that for $A=0.8$ and a
reasonable range of values of $B$ all the eigenvalues were very close to
those found at zeroth order.  Results after four iterations, for various
values of $B$, are given in Table II.

\newsec{The Magnetic Deformation Exponent}

Thus far we have only considered $Z_2$-symmetric perturbations about
the fixed point. It is also interesting to consider
$Z_2$-antisymmetric deviations from the fixed point, since these yield
another critical exponent $\l_H$. In
the analogous Ising system, such perturbations measure the response of
the system to a weak external magnetic field.

To find these antisymmetric eigenvectors to eq. \lineq, we use as
initial conditions $g_n(0)=1, g'_n(0)=0, h_n(0)=0$ and
$h_n'(0)=h_0$ very small.  This does not give an independent
critical exponent for
the theory, but it does allow us to verify for $d=3$
the `scaling relation' \itdr
\eqn\scalrel{\l_H=1+{d\over2}-{\eta_*\over2}\ ,}
giving us a nontrivial consistency check on our results for $\eta_*$.  At order
$p^0$ and at order $p^2$, with $A=0.8$ and several values of $B$, we
have calculated $\l_H$, and have confirmed numerically that the
relation \scalrel\ is always satisfied.

Indeed it is possible to prove \scalrel\ rigorously for any
$d$, any scheme, and at leading or next to leading order in the
derivative expansion: it is straightforward to verify that
\eqn\exacteigen{(g_n(x),h_n(x))=(f'_*(x)-(1-\smallfrac{\eta}
{2}),z'_*(x))}
is an exact eigenvector of eq. \lineq, with eigenvalue
$\l_H$ given by eq. \scalrel.

Incidentally,
\eqn\exacteigentwo{(g_n(x),h_n(x))=
(f'_*(x),z'_*(x))}
 is also an exact eigenvector, with eigenvalue $\l_I
=-1+\smallfrac{d}{2}+\smallfrac{\eta}{2}$.  At $d=3$, there are thus
three relevant eigenvalues: $\l_T\equiv 1/\nu$, $\l_H$ and $\l_I$.  Just as
$\l_H$ and $\l_T$ correspond to perturbations proportional to the
operators $\vf$ and $\vf^2$, respectively, $\l_I$ corresponds to
perturbations containing the operator $\vf^3$, but proportional to the
equations of motion.

\topinsert\hfil
$$\vbox{\tabskip=0pt \offinterlineskip

\halign to 380pt{\strut#& \vrule#\tabskip=1em plus2em&
   \hfil#& \vrule#&
   \hfil#& \vrule#& \hfil#& \vrule#&
   \hfil#& \vrule#\tabskip=0pt\cr
\noalign{\hrule}
& &\hidewidth \hidewidth
& &\hidewidth $\eta$ \hidewidth
& &\hidewidth $\nu$ \hidewidth
& &\hidewidth $\omega$ \hidewidth &
 \cr \noalign{\hrule}
 && {\rm LO}       && {\rm 0}     & & $0.649$
               & & $0.66$         & \cr\noalign{\hrule}
&& {\rm NLO }
                   && $0.019$ - $0.056$
                                & & $0.616$ - $0.637$
   & & $0.70$ - $0.85$  &\cr\noalign{\hrule}
 && {\rm NLOS}      && $0.035$  & & $0.627$
               & & $0.76$  & \cr\noalign{\hrule}
 && {\rm Ref.\itdr}&& $0.030$ - $0.040$ & & $0.630$ -  $0.631$
               & & $0.75$ -  $0.85$  &\cr\noalign{\hrule}
\noalign{\smallskip}}}$$
\hfil
\centerline{\vbox{\hsize= 300pt \noindent\footnotefont
Table~III: Critical exponents at leading order (LO), next-to-leading
order for $B$ in the range .25 - .75 (NLO) , next-to-leading order
with the scheme dependence fixed to give
$\eta$ correctly (NLOS), and finally the average value obtained from
other methods \itdr.}} \bigskip
\endinsert

\newsec{Conclusions}

Like perturbation theory, the derivative expansion to any finite order
represents a truncation of the renormalization group.  We find it
interesting that our results for anomalous dimensions are scheme
independent (or, more precisely, cutoff independent) to first
nontrivial order and scheme dependent beyond that.

Our results for the leading critical exponents of a scalar field theory
with $Z_2$ symmetry in three dimensions at the Wilson fixed point are
summarized in Table~III, together with values obtained from different
methods \itdr (resummed perturbation series, $\epsilon$-expansions,
lattice simulations, etc.).  At leading order our (scheme independent)
equation seems to do remarkably well; essentially this is because the
true value of $\eta$ is very small.  At next-to-leading order the
results improve in the right direction (and are consistent with those in
\refs{\timone,\tetwet}), but scheme dependence makes it impossible to
give a precise estimate of the errors at this order. Our
attempts to fix the scheme by some sort of `minimal sensitivity'
criterion were only partially successful, because $\eta$ depends
approximately linearly on the scheme parameter $B$.
A next-to-next-to-leading order computation might give results with
a weaker scheme dependence, allowing perhaps for a determination of
$\eta$ and thus a sensible estimation of the errors incurred in other
quantities also. However, at present scheme dependence severely limits
the accuracy of the technique, in particular for quantities which
depend sensitively on $\eta$.  The scheme ambiguity
of our results may however be significantly reduced if we simply
choose a scheme which gives
the `known' value for $\eta$: our results for $\nu$ and $\omega$ are
then both competitive and consistent with the `known' results for these
exponents.

We also discussed two other relevant eigenvalues, corresponding to asymmetric
perturbations about the fixed point.  We showed that the scaling
relations relating these relevant eigenvalues to the other critical
exponents are satisfied exactly by our truncated set of equations,
and thus give a useful consistency check on the numerical calculations.

It should be possible in the future to use these truncated \ERG\
equations to compute other quantities of physical interest (and in
particular Green's functions), in a wider range of models.  In
particular it would be possible \ref\mag{M.~Maggiore,
\ZP\vyp{C41}{1989}{687}\semi
U.~Ellwanger and L.~Vergara, \NP\vyp{B398}{1993}{52}\semi
T.E.~Clark, B.~Haeri and S.T.~Love, \NP\vyp{B402}{1993}{628}.}
to include (chiral) fermions,
and thus study (for example) the electroweak sector of the standard
model (with gauge couplings suppressed).

\bigskip
\vbox{
{\bf Acknowledgments}\nobreak\smallskip\nobreak
It is a pleasure to thank Poul Damgaard for many helpful
discussions, and also Tim Morris and Roberto Percacci for their
careful readings of the final manuscript.
This work is supported in part by funds provided by AEN 90-0033
Grant (Spain), by M.E.C (Spain) and by CIRIT (Generalitat de Catalunya).
}
\listrefs
\listfigs
\end